\documentclass[generic,preprint]{imsart}

\RequirePackage[OT1]{fontenc}
\RequirePackage{amsthm,amsmath}
\RequirePackage[numbers]{natbib}
\usepackage{graphicx}

\startlocaldefs
\numberwithin{equation}{section}
\theoremstyle{plain}

\endlocaldefs

\begin{document}

\begin{frontmatter}
\title{Set-based differential covariance testing for high-throughput data}
\runtitle{Set-based covariance testing}

\begin{aug}
\author{\fnms{Yi-Hui} \snm{Zhou*}\\{yihui\_zhou@ncsu.edu}}

\address{Bioinformatics Research Center and Department of Biological Sciences\\
North Carolina State University\\
Campus Box 7566\\
Raleigh, NC 27695\\}


\end{aug}

\begin{abstract}
The problem of detecting changes in covariance for a single pair of features has been studied in some detail, but may be limited in importance or general applicability. In contrast, testing equality of covariance matrices of a {\it set} of features may offer increased power and interpretability. Such approaches have received increasing attention in recent years, especially in the context of high-dimensional testing. These approaches have been limited to the two-sample problem and involve varying assumptions on the number of features $p$ vs. the sample size $n$. In addition, there has been little discussion of the motivating principles underlying various choices of statistic, and no general approaches to test association of covariances with a continuous outcome.   We propose a uniform framework to test association of covariance matrices with an experimental variable, whether discrete or continuous. We describe four different summary statistics, to ensure power and flexibility under various settings, including a new ``connectivity" statistic that is sensitive to changes in overall covariance magnitude.  The approach is not limited by the data dimensions, and is applicable to situations where $p >> n$. For several statistics we obtain asymptotic $p$-values under relatively mild conditions. For the two-sample special case, we show that the proposed statistics are permutationally equivalent or similar to existing proposed statistics. We demonstrate the power and utility of our approaches via simulation and analysis of real data. 

\end{abstract}



\end{frontmatter}

\section{Introduction}
Tests of changes in correlations or covariances have received considerable attention in areas such as genomics \citep{mckenzie2016dgca} and finance \citep{isogai2016building}. The liquid association method 
 \citep{li2002genome} was developed to decribe ternary relationships among genes, but the underlying motivation is based on the concept of differential covariance. For tests of biological systems, ensemble  tests of covariance matrices for multiple genes in a proposed network or pathway can provide insight that might not be apparent in tests of individual gene pairs \citep{yuan2016network}. Classical likelihood approaches require the sample size $n$ to be large compared to the number of features $p$ \citep{john1971some, anderson1962introduction}. These tests have been largely limited to the two-sample problem 
of testing equality of $p\times p$ covariance matrices $H_0: \Sigma_1=\Sigma_2$ based on samples of sizes $n_1$ and $n_2$, where $n_1+n_2=n$.  In settings where $p>min\{n_1,n_2\}$, likelihood ratio testing may perform poorly or be undefined.  

Recently a number of investigators have reconsidered the two-sample problem where $p>n$.
Li and Chen
(\cite{li2012two}) derived an approximately standard normal statistic for the Frobenius norm of differences in the two $p\times p$ sample covariance matrices, with considerable attention to sources of bias when $p$ is large.
 \cite{cai2013two} proposed a maximum standardized difference statistic between two sample
covariances, with testing based on an extreme value approximation.  
The two approaches were designed for very different alternatives, ranging from modest but widespread differences in the two sample covariance matrices (\cite{li2012two}) to large differences in a very few covariance elements (\cite{cai2013two}).

The two-sample problem can be viewed as an ``association" of the covariance matrix with a binary
group indicator.  More generally, the investigator may be interested in trend association of covariance with an experimental variable $y$ that might be multi-level, or on a continuous scale. To our knowledge, no general method is available with the requisite flexibility, without restrictive parametric requirements or assumptions of the feature size $p$ relative to $n$. Moreover, existing methods have been published in isolation, providing little opportunity to consider power characteristics for various types of alternatives. 

 In this manuscript, we propose four different statistics to test changes in a covariance matrix of $p$ features, when it is anticipated that a change in $y$ will result in (1) a directional change in elements of the covariance matrix; (2) a non-directional change in covariance; (3) a change in the overall magnitude of covariances; or (4) a large change in one or a few elements of the covariance matrix. 
In contrast to almost all of the relevant literature, the statistics apply naturally whether $y$ is continuous or discrete. Asymptotic results for most of the statistics are derived, to provide computationally efficient $p$-values. For small sample sizes, permutation can be used to ensure control of type I error. 

This paper is organized as follows. In Section \ref{method}, we use a motivational example dataset
to introduce the method and test statistics. Section \ref{theorem} establishes permutation equivalence between some of the proposed statistics and  existing methods for the two-sample problem. In Section \ref{simulation}, we compare the proposed  statistics with existing methods, in terms of type I error and power. Several different simulation settings are presented for the two-sample problem, comparing our statistics to existing methods.  In addition, we compare our proposed methods in the setting with continuous $y$. 

\section{Methods}\label{method}
\subsection{Notation}
Let $X$ be the $p \times n$ data matrix consisting of elements $x_{ik}$, and
and $y$ the $n$-vector of clinical/experimental data. The sample mean and variance of a vector follow standard notation, e.g. $\bar{y}$ and $s_y^2$.  
The $i$th row and $k$th column of $X$ are denoted $x_{i.}$ and  $x_{.k}$, and each column is assumed to have a population
$p$-mean of zero.  Random variables are capitalized (e.g, random $Y_k$ vs. observed $y_k$).
We denote the $p\times p$ covariance of $X$, which may depend on $y_k$, as $\Sigma_{y_k}$. The zero-mean assumption is implicit in most covariance tests, following an intent that the test statistics be sensitive only to changes in covariance.  
For a subset of samples $\omega$ with at least 2 samples, the sample covariance is $\widehat{\Sigma}_\omega= X_\omega X_\omega^T/n_\omega$.
A single $i, j$ element is 
$\hat\sigma_{i j, \omega}=(\sum_{k\in\omega}x_{ik}x_{j k})/n_\omega$.
We use $\xi$ to denote the operator that sums all elements of a matrix, and the superscript ``${\circ k}$" to denote the element-wise exponent of a matrix to power $k$.  

We motivate the concepts and potential utility of our methods by considering gene expression data on  kidney transplant tissue \citep{modena2016gene}, in which those with acute rejection ($y=1, n_1=54$) were compared with normal outcomes ($y=2, n_2=99$).  To identify biological pathways (gene sets) of interest in the comparisons, groups of genes for KEGG and Gene Ontology were identified, so that for each test, $p$ represents the number of genes in the pathway.  To identify pathways of greatest interest, we aggregate over all gene pairs in each pathway, using statistics as described below.

A  ``summation" statistic $S=\xi(\widehat{\Sigma}_1-\widehat{\Sigma}_2)$ is sensitive to covariance changes in the same direction, although testing would of course  be two-sided.  A statistic $Q=\xi \bigl((\widehat{\Sigma}_1-\widehat{\Sigma}_2)^{\circ 2}\bigr)$ is sensitive to covariance changes in either direction that might otherwise cancel in $S$.   A statistic $C=\xi (\widehat{\Sigma}_1^{\circ 2}-\widehat{\Sigma}_2^{\circ 2})$ is intended to provide additional weight for genes with large changes in covariance magnitude. Finally, a statistic $M$ finds the maximum $(\hat\sigma_{i j, 1}-\hat\sigma_{i j, 2})^2$ (i.e. elements of the matrix used for $Q$), appropriately scaled by a standard error as described further below. 

\begin{figure}[t]
\begin{center}
\includegraphics[width=17.5cm]{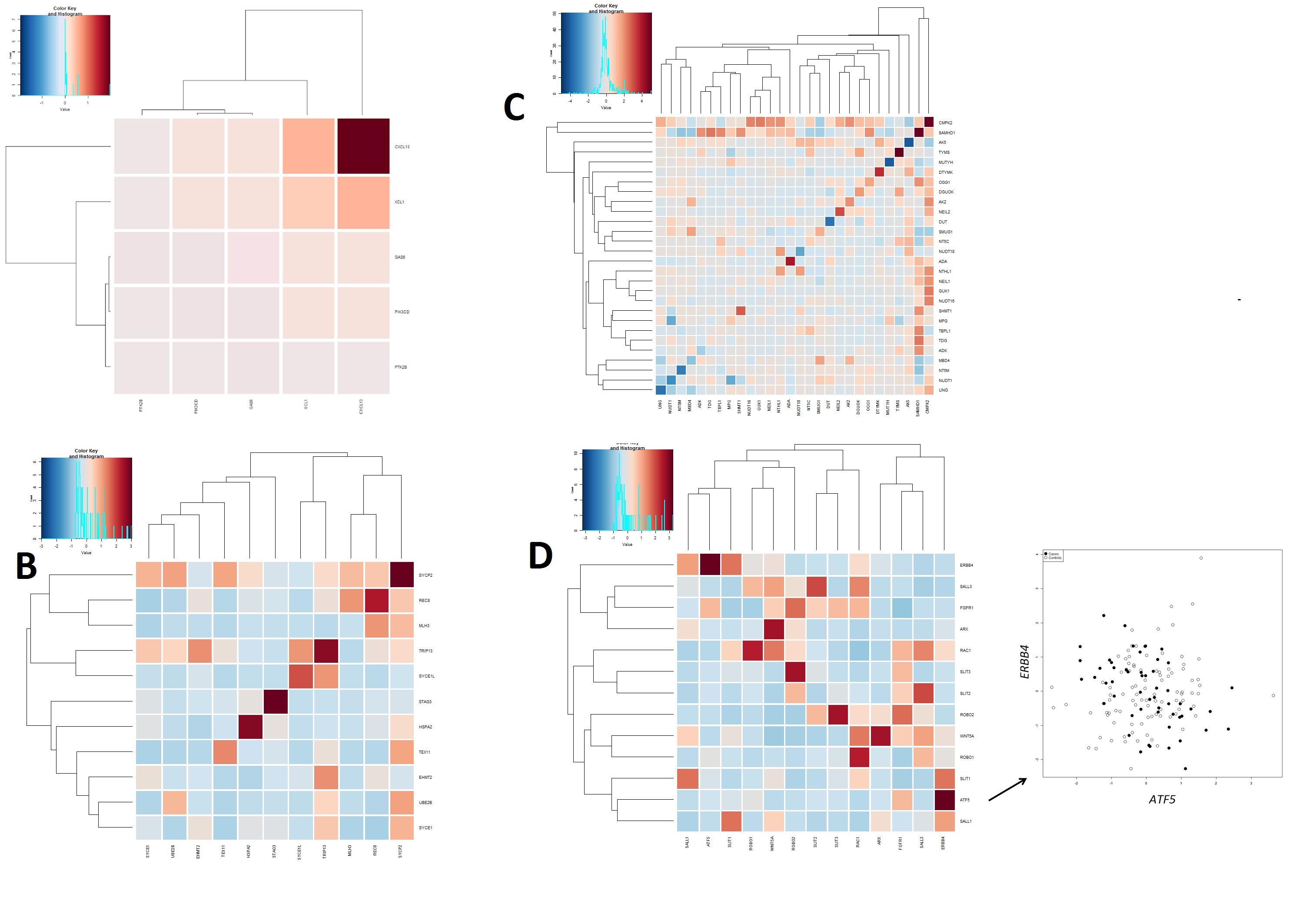}
\caption{Illustration of the 4 statistics in the kidney transplant data, using gene set analysis of gene expression in those with acute rejection ($n_1=54$) vs. normals ($n_2=99$). All panels except for panel (A) were zero-centered to better illustrate the covariance changes. 
	(A) Heatmap of $\widehat{\Sigma}_1-\widehat{\Sigma}_2$ for GO:0035754, the most significant pathway for $S$.
	(B) Heatmap of $(\widehat{\Sigma}_1-\widehat{\Sigma}_2)^{\circ 2}$ for GO:0070193, the most significant pathway for $Q$.
	(C) Heatmap of $(\widehat{\Sigma}_1^{\circ 2}-\widehat{\Sigma}_2)^{\circ 2}$ for GO:0009394, the most significant pathway for $C$.
	(D) Heatmap of $(\widehat{\Sigma}_1^{\circ 2}-\widehat{\Sigma}_2)^{\circ 2}$ for GO:0021889, the most significant pathway for $M$. The inset shows the covariance in acute rejection vs. controls for $ATF5$ and $ERBB4$. }
\label{4panel}
\end{center}
\end{figure}

Methods for obtaining $p$-values will be described, but first we illustrate by showing the most significant pathways for each statistic in Figure \ref{4panel}. Each heatmap depicts the matrix corresponding to the statistic (e.g. for $S$ it depicts $\widehat{\Sigma}_1-\widehat{\Sigma}_2$). The most significant pathways for each statistic are as follows, with multiple comparison false discovery $q$-values: GO:0035754 B cell chemotaxis (for $S$, $q=1.4\times 10^{-6}$),  GO:0070193 synaptonemal complex organization (for $Q$, $q=0.03$), and  GO:0009394 2'-deoxyribonucleotide metabolic process (for $C$, $q=4.2\times 10^{-18}$).

Interpretation of $M$ as a gene-set statistic is less straightforward, because it is based on the maximum covariance difference of feature pairs, and thus may not be viewed as a ``pathway" finding. Nonetheless, as an illustration, we show the results for the most significant $M$ statistic for GO:0021889 olfactory bulb interneuron differentiation (Figure \ref{4panel}D, $p$-value=0.0005, $q$ n.s.). To best illustrate changes in correlation rather than changes in variance, for this statistic we row-scaled the data to have variance 1 for each gene.
The gene pair \{{\it ATF5, ERRB4}\} shows the most significant change, with a high negative correlation in the AR group and little correlation in the normal group.
{\it ATF5} has been associated with transplant rejection in multiple organ systems \citep{morgun2006molecular}. There is little literature on {\it ERBB4} and transplant rejection, but the gene has been associated with kidney nephropathy  \citep{sandholm2012new} and thought to be protective of polycystic kidney disease in a mouse model \citep{zeng2014deletion}.

\subsection{A conceptual trend model}

To motivate our statistics, we adopt a conceptual trend model for the covariance dependence of $X$ on $y$: $\Sigma_y=\beta_0+\beta_1 y$ for $p\times p$ matrices $\beta_0$, $\beta_1$. Thus for sample $k$, 
according to our assumptions $E(X_{ik}|y_k)=0$ for each $i$, and
${\rm cov}(X_{ik} X_{j k}|y_k)=E(X_{ik} X_{j k}|y_k)= \beta_{0,i,j}+\beta_{1,ij}y_k$ for the $i$th and $j$th feature.
Letting $z_{ij k}=x_{ik}x_{j k}$, the model immediately suggests linear regression of $z$ on $y$, for which the least-squares slope solution is $\widehat{\beta}_{1,i j}=(\Sigma_k z_k y_k/n-\overline{z}\overline{y})/(s^2_y \frac{n-1}{n})$.
Although the trend assumption is simple, $\hat{\beta_1}$ can be viewed as an approximate score statistic for models $E(Z_{ij})=\eta_{0,i,j}+\eta_{1,ij} f(y_k)$ for strictly monotone
smooth $f$, and thus locally powerful for detecting departures from the null $\beta_{1,ij}=0$.
We make two further observations: (1) we do not consider $\beta_{0, i j}$ to be of interest for detecting covariance changes, and (2) linear rescaling of $y$ will not meaningfully change our results, because
it results in constant changes in the proposed statistics. Thus without loss of generality we  assume $\bar{y}=\sum_k y_k/n=0$, so $\widehat\beta_{1, i j}=\frac{1}{n s_y^2} \sum_k x_{ik} x_{j k} y_k=\frac{1}{n s_y^2} \sum_k z_{ij,k} y_k$.
These least squares solutions are not intended to be used directly, but serve to motivate global test statistics described below. 

\subsection{A summation statistic}
To effectively measure the covariance changes, we  propose $S=\sum_i \sum_{j} \hat\beta_{1, i j}$ as a {\it summation} statistic to detect global changes in covariances that are  concordantly associated with the experimental variable $y$ (i.e., in the same direction).
A simplification for $S$ is 
\begin{eqnarray*}
S& =& \sum_i \sum_{j}
  \sum_k x_{ik} x_{j k} y_k= \sum_k y_k \sum_i x_{ik} \sum_{j} x_{j k}= \sum_k y_k (\sum_i x_{ik})^2 = \sum_k w_k y_k =y^T w 
\end{eqnarray*}
for $w_k=(\sum_i x_{ik})^2$. 
In datasets where the null for $S$ can be rejected,
the value $w_k$ represents a natural ``risk score" for sample $k$, with extreme $w$ values corresponding to extreme $y$.

Although the initial motivation for $S$ was based on $p\times p$ covariance terms, the restated statistic is ultimately based on an inner product of $n$-vectors, and thus we may use a large-sample normal approximation for rescaled $S$ to obtain $p$-values (Appendix). To compute somewhat more accurate $p$-values for small $n$, for $S$ (and $C$ below) we apply the moment-correlated correlation (MCC) method of \cite{zhou2015hypothesis}, which uses the skewness and excess kurtosis to refine the $p$-values in comparison to a normal approximation. However, the normal limiting approximation still applies to MCC, as these additional moments tend to zero.

The proof, which is a standard triangular array version of the central limit theorem, allows for
dependence of $p$ on $n$ and is in the Appendix.
 Verifying the conditions requires a model for $X$ with increasing $n$, but will be satisfied for a wide variety of `typical' assumptions. For example, previous work has often assumed multivariate normality of $X$ and either a binary or normal $Y$, for which the conditions can be shown to be satisfied (see Appendix) under very general conditions.

\subsection{A quadratic form statistic}

In contrast,  $Q=\sum_i \sum_{j} \widehat\beta_{1, i j}^2$ is sensitive to changes that are not directionally concordant.  Similar to $S$, $Q$ can also be represented by $n$-vectors and $n\times n$ matrices.
\begin{eqnarray*} 
Q & = & \sum_i \sum_{j} (\sum_k x_{ik} x_{j k} y_k)^2 =  \sum_i \sum_{j} \sum_k \sum_{l} x_{ik} x_{j k}  x_{i l} x_{j l} y_k y_{l} \\
&=&\sum_k\sum_{l} y_k y_{l} \sum_i x_{ik} x_{i l} \sum_{l} x_{jk} x_{j l} =\sum_k\sum_{l} y_k y_{l} a_{k l}
\end{eqnarray*}
 where $a_{kl}=(\sum_{i} x_{ik} x_{i l})^2$.  The matrix with elements $a_{k l}$ can be simplified to $A=(X^T X)^{\circ 2}$. Finally, we have the quadratic form $Q=y^T A y$.

 The nature of $Q$ makes it difficult derive a risk-score analogue, and also difficult to justify closed-form limiting approximations to its null distribution. An exception is in extreme cases, such as dominance of a single eigenvalue in $A$ (approximately chisquare), or with a large number of eigenvalues of similar magnitude (approximately normal). 
As explored in \cite{zhou2013space}, for small to moderate sample sizes, a weighted beta approximation can be more accurate than standard approximations for sums of independent chisquare distributions. However, the procedure can be somewhat computationally intensive, and here we opt for direct permutation of $y$ to obtain $p$-values, for general $A$.  For certain special cases, solutions for the first four permutation moments \citep{zhou2013space} may be used to obtain $p$-values, but require restrictive assumptions on the form of $A$.

\subsection{A connectivity statistic}
Each element $a_{k l}$ of $A$ has the form of a squared inner product between samples $k$, $l$, and so $b_k=\sum_{l} a_{k, l}$ reflects broad-scale association (a ``connectivity index") of sample $k$ with remaining samples.  Accordingly, we propose the {\it connectivity} statistic $C=y^T b$ for risk scores $\{b_k\}$ to reflect correlation between $y$ and the connectivity index.  Correlations between samples are ultimately driven by correlation between features, and $C$ reflects the tendency for the aggregate magnitude of feature-feature correlations to be associated with $y$, which is quite different from the type of alternative envisioned for $S$ and $Q$. Theorem 1 also applies to $C$, with the theorem conditions applied to $\{B_k^2\}$, so that $p$-values are obtained in the same manner as with $S$.


\subsection{A maximum statistic}

Our fourth statistic was inspired by 
\cite{cai2013two}, who devised a test for the maximum element difference, scaled by an appropriate standard error, for sample covariance
matrices in the two-sample problem.  
Again, we wish to generalize the statistic, 
and define
$M_{ij}= (n-1)r_{ij}^2$,
where $r_{ij}$ is the Pearson correlation between $y$ and $z_{ij.}$.
Defining $\{i^\prime,j^\prime\}={\rm argmax}_{i,j}M_{ij}$,
we propose the {\it maximum} statistic $M=M_{i^\prime,j^\prime}$,
with risk score $z_{i^\prime,j^\prime}$. 

Approximate $p$-values follow from the theorem of \cite{cai2013two} for their maximum statistic, analogous to ours, using an extreme value approximation for $\chi_1^2$ variates.
 Beyond standard assumptions that elements of $X$ and $Y$ have appropriate tail behavior (e.g. sub-Gaussian), there are modest restrictions on $\Sigma$, and $p$ should grow slower than $n^5$. 
With these assumptions, approximate $p$-values are obtained using $P(M - 4 \log p +\log \log p \leq t ) \approx \exp(-\frac{1}{\sqrt{8\pi}} \exp(-\frac{t}{2}))$.

\subsection{Permutation testing}
Permutation testing can be useful, both as a means of performing small sample analysis, and in informing interpretation of our statistics, as we show in the next subsection.
Letting $\Pi$ denote a random permutation of $n$ elements from among the $n!$ possibilities (realized value $\pi$), the statistics for permutation $\pi$ are  
$S_{\pi}=y_\pi^T w$, $Q_\pi=y_\pi^T A y_\pi$, $C_\pi=y_\pi^T b$, and $M_\pi$ (which requires
computation of the $\hat\beta_1$ values and standard errors for each permutation).
$S$ and $C$ are subjected to two-sided testing, with $p$-values based on both right and left tails, while $Q$ and $M$ are one-tailed, rejecting for large values.  For example, with $H$ random permutations
and $\pi[h]$ denoting the $h$th permutation, the empirical $p$-value for $S$ is
$p_S=\sum_{h=1}^H I[|S_{\pi[h]}|\ge |S_{observed}|]/H$, while the $p$-value for $Q$ is
$p_Q=\sum_{h=1}^H I[Q_{\pi[h]}\ge Q_{observed}]/H$.

The null hypothesis is that the relationships of columns of $X$ to the elements of $y$ are 
exchangeable (\cite{good2002extensions}), which holds if $X$ and $y$ are drawn from independent
distributions. 
A primary advantage of permutation testing is that, aside from slight issues due to discreteness or tied outcomes, type I error rates are controlled without requiring parametric assumptions
(\cite{zhou2015hypothesis}). This property is especially important for covariance association testing, enabling implementation for data of any size $p$ and $n$.

\subsection{Two group comparisons and permutation equivalence}\label{theorem}

The following result ties our proposed statistics to two intuitive statistics, including those proposed by others, in comparing two sample covariance matrices.

\medskip
\noindent \textbf{Result 1.} Let $\omega_1$ and $\omega_2$ be the indexes for samples in groups 1 and 2, respectively, and the subscripts 1 and 2 will be used for simplicity. We assign the experimental variable
$y_k = \frac{1}{n_1}$ if $k\in\omega_1$, and $y_k=\frac{-1}{n_2}$ if $k\in\omega_2$. Then

(i) The directional statistic $S$ is equivalent to $\xi(\widehat{\Sigma}_1-\widehat{\Sigma}_2)$,

(ii) The non-directional statistic $Q$ is equivalent to  
$\xi \bigl((\widehat{\Sigma}_1-\widehat{\Sigma}_2)^{\circ 2}\bigr)$.

\noindent
{\it Proof.} We have covariance element differences 
\[\hat\sigma_{i j,1}-\hat\sigma_{i j,2}=\frac{\sum_{k\in\omega_1} x_{ik}x_{j k}}{n_1}-\frac{\sum_{k\in\omega_2} x_{ik}x_{jk}}{n_2}=\sum_k x_{ik} x_{jk} y_k=\hat\beta_{1,i j}.
\]
Summing over the $p\times p$ elements we have $\xi(\widehat{\Sigma}_1-\widehat{\Sigma}_2)=\sum_i\sum_{j} \hat\beta_{1, i j}$, and $\xi\bigl((\widehat{\Sigma}_1-\widehat{\Sigma}_2)^{\circ 2}\bigr)=\sum_i\sum_{j} \hat\beta_{1, i j}^2$.

\medskip
We note that $\xi\bigl((\widehat{\Sigma}_1-\widehat{\Sigma}_2)^{\circ 2}\bigr)$ is essentially the Frobenius norm statistic proposed by \cite{li2012two}, except that the authors employed various bias corrections (because $E(\hat\Sigma)\ne \Sigma)$
 to construct their statistic.  When using permutation, such corrections are unnecessary, because the observed and permuted values are subject to the same bias.  Additionally, two different statistics
will provide permutation $p$-values that are identical if the statistics are {\it permutationally equivalent} as defined in section 2.4 of \cite{pesarin2010permutation}.  Thus it is immaterial that the Frobenius norm 
involves a square root not used in the statistic shown here (\cite{golub2012matrix} pg. 55). Moreover, there is no need for standard error estimation (e.g. as employed by \cite{li2012two} to compute an approximately $N(0,1)$ statistic).

\begin{figure}[t]
	\begin{center}
		\includegraphics[width=15cm]{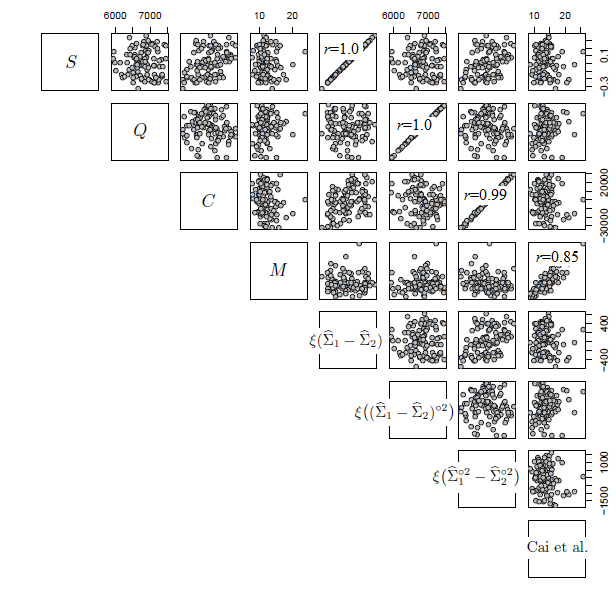}
		\caption{Comparison of the four proposed statistics to various existing statistics for the two-sample problem, for a single simulated dataset and 100 permutations. Pearson correlations illustrate the exact and approximate correspondence of some pairs of statistics.}
		\label{perm}
	\end{center}
\end{figure}

Figure \ref{perm} shows the results from 100 random permutations of $y$ for the two sample problem with $n_1=n_2=20$, $p=50$. A single $X$ was generated using the null version of Model 2 described in the next section, but the qualitative results hold regardless of the choice of $X$.
 As we showed above, $S$ and $Q$ are equivalent to $\xi(\widehat{\Sigma}_1-\widehat{\Sigma}_2)$ and $\xi\bigl((\widehat{\Sigma}_1-\widehat{\Sigma}_2)^{\circ 2}\bigr)$ respectively. Under the permutations, $C$ has high Pearson correlation over permutations with 
$\xi\bigl(\widehat{\Sigma}^{\circ 2}_1-\widehat{\Sigma}^{\circ 2}_2\bigr)=
\xi(\widehat{\Sigma}^{\circ 2}_1)-\xi(\widehat{\Sigma}^{\circ 2}_2)
$, supporting the perspective that $C$ reflects a contrast in the overall magnitude of covariances. Finally, our $M$ is correlated under permutation with the statistic from \cite{cai2013two}, although they differ modestly due to differences in the standard errors used.  

This permutation example underscores the correspondence between our statistics and those that seem ``natural" for the two-sample problem, but we emphasize that our statistics apply for either discrete or continuous $y$.

\section{Type I error and Power}\label{simulation}
We use the analyic $p$-value approximations for our statistics, except for $Q$, which uses permutation. For the two sample problem, we start by examining the type I error and power characteristics for the proposed and existing statistics, and follow with power analyses for the proposed statistics for some settings with continuous $y$.

\subsection{Two sample comparisons} \label{twosample}

Initial comparisons follow the simulation settings from \cite{li2012two}, for which feature covariances were described using auto-regressive notation. More compactly, we describe their simulation settings in terms of the covariance matrices.

\subsubsection{Simulation Model 1 (type I error)}

We assume the first population $X_1$ $\sim N(0, \Sigma_1)$; while the second population $X_2$ $\sim N(0, \Sigma_2)$, where 

$\Sigma_{1ij} = \begin{cases} 1+\theta^2_1, & \mbox{if } i\mbox{  $=j$} \\ \theta_1, & \mbox{if } i\mbox{ $= j+1  or  j-1$} \\
0, & \mbox{if } i \mbox { $ \neq j-1,j,j+1$} \end{cases}$,
$\Sigma_{2ij} = \begin{cases} 1+\theta^2_1+\theta^2_2, & \mbox{if } i\mbox{ $ =j$} \\ \theta_1(1+\theta_2), & \mbox{if } i\mbox{ =$ j+1 \ or  j-1$} \\
0, & \mbox{if } i\mbox { $ \neq j-1,j,j+1$}\end{cases}$.

\noindent The difference between the two covariance matrices is 
$$\Sigma_{2ij}-\Sigma_{1ij} = \begin{cases} \theta^2_2, & \mbox{if } i\mbox{ $=j$} \\ \theta_1 \theta_2, & \mbox{if } i\mbox{ $= j+1 \ or  j-1$} \\
0, & \mbox{if } i\mbox { $ \neq j-1,j,j+1$}\end{cases}.$$ To assess type I error, we set $\theta_2=0$, which implies the null $\Sigma_{1}-\Sigma_{2}=0$. We show results for
$n_1=n_2=\{20,50,80,100\}$ and feature dimension $p=\{32,64,128,256,512,700\}$. The number of simulations was 1000 for each setting. Here and in subsequent tables the MCC parametric method was used to obtain $p$-values for $S$ and $C$, and 1000 permutations for $Q$. Although the parametric method works well for $M$ for moderate sample size, to ensure robustness for this statistic for the entire range of simulations we report permutation-based $p$-values.

\begin{figure}[t]
	\begin{center}
		\includegraphics[width=15cm]{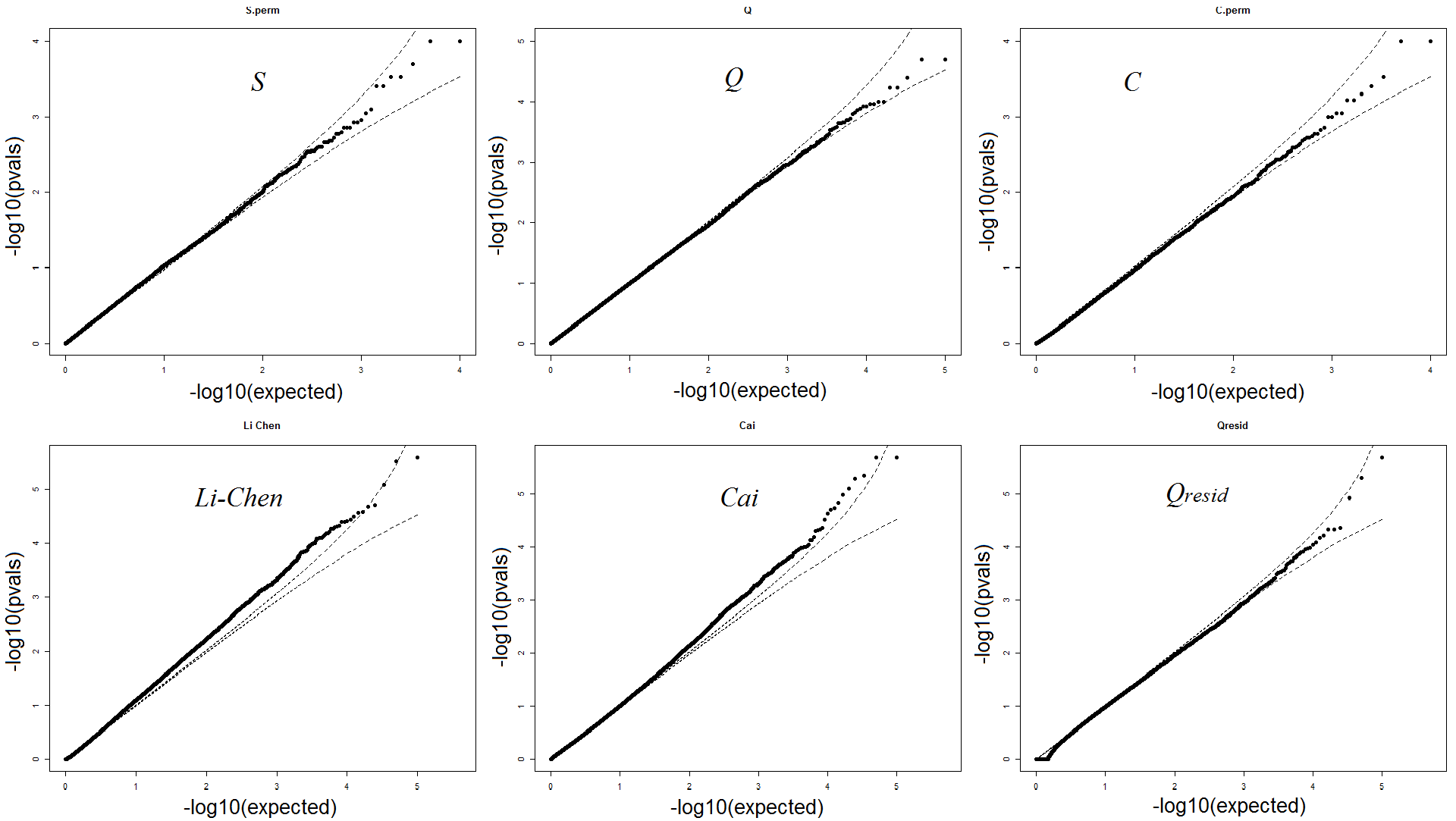}
		\caption{QQplots for several of the proposed methods, as well as existing methods, for the null two-sample problem of Simulation Model 1, $p=50$, $n_1=n_2=20$.}
		\label{Fig:qqplots}
	\end{center}
\end{figure}

Table \ref{tab:typeInormal} shows that for this multivariate normal model, most methods perform well and control type I error.
The $Cai$ (\cite{cai2013two}) method is noticeably anticonservative for $\alpha=0.05$ for the smaller sample size ($n_1=n_2=20$), and more so as $p$ increases.  For the setting with $n_1=n_2=20$, $p=50$, 100,000 simulations were performed to provide greater insight into tail behavior (Figure \ref{Fig:qqplots}). $p$-values for $Q$ perform well, which is sensible, as the permutation null holds.  In addition, we also show good results for a ``residualized" $Q$ (lower right panel), in which each row of $X$ is residualized using simple linear regression for the effect of $y$. The rationale for such an approach might be to ensure that the test statistic is sensitive to changes in covariance only, not to any linear association with $y$.
Here the residualization is also performed inside the permutation loop.

\subsubsection{Simulation Model 2 (type I error)}

Here we follow the previous Simulation Model but with skewed data elements.  Specifically, let $G(w;4,0.5)$ denote the gamma distribution function with shape parameter 4 and scale 0.5 evaluated at $w$.  Then if $W\sim G$, $X=W-2$ has mean zero and variance 1, i.e. follows a centered gamma.  The elements of $X_1$ and $X_2$ are drawn as shown above, following the same null covariance structure that was used in Simulation Model 1.

Here the $Cai$ approach in \cite{cai2013two} becomes conservative, both with increasing sample size and  feature size (Table \ref{tab:typeIgamma}).  The $Li-Chen$ method is anti-conservative, but the type I error becomes closer to nominal as the sample size and feature size increase.
%
As expected, our proposed permutation methods are very robust in controlling type I error for all $n$ and $p$, as the skewness in data elements does not violate the exchangeability property.

\subsubsection{Simulation Model 3 (power)}

For power comparisions, we return to the multivariate normal data elements. We use Simulation Model 1, but with covariance matrices determined by $\theta_1=2, \theta_2=1$ (one of the simulation models also used by \cite{li2012two} and summarized in their Table 4). Although this simulation model was used by \cite{li2012two} to support their proposed statistic, our proposed $C$ has consistently highest power for all the $n, p$ settings.  The $Li-Chen$ statistic shows power slightly higher than that of $Q$, even though
they both are based on the Frobenius norm. We speculate that the reason is related to the fact that permutation testing is conditional on the observed data, and the power difference nearly disappears at the larger sample sizes.  It is perhaps a bit surprising that $S$ is less powerful than $Q$, as the covariance differences are directional. However, the squared terms in $Q$ also may effectively act to reduce noise, and we later show situations in which $S$ is more powerful.  The $Cai$ and $M$ statistics show the lowest power, as they use only the most extreme covariance difference element, and do not aggregate over the large number of covariance difference elements.

\subsection{Simulations with a continuous $y$} \label{cont}

\subsubsection{Simulation Model 4}
\begin{figure}[t]
\begin{center}
\includegraphics[width=8cm]{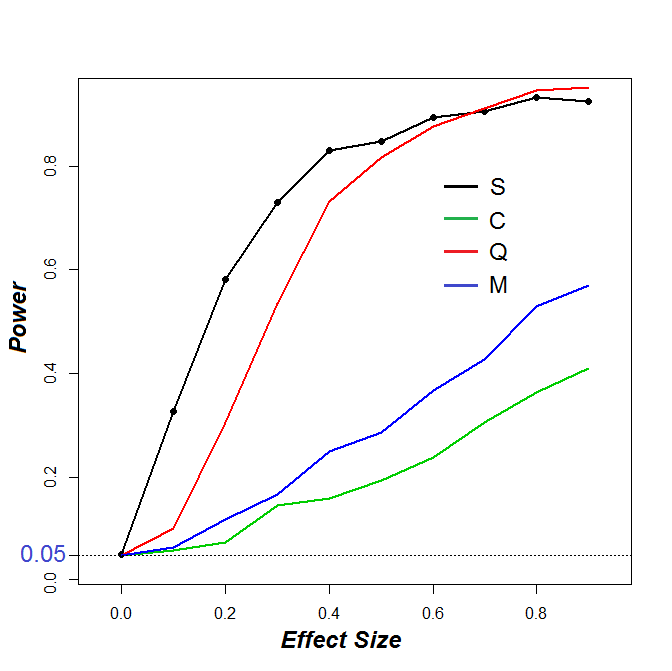}
\caption{Power comparision among $S$, $Q$, $C$, $M$ for Simulation Model 4. The dashed line at $\alpha=0.05$ indicates that all the proposed methods control type I error well under the null ($\rho=0$). The effect size $\rho$ ranges from $0$ to $0.8$. }
\label{Fig:power_conty}
\end{center}
\end{figure}

For this simulation model, values in $y$ are drawn iid $N(0,1)$ in each simulation, and converted to
the re-scaled experimental variable $y^*=\frac{y-min(y)}{max(y)-min(y)}  \in [0,1]$.
$X$ is drawn as multivariate $N(0,\Sigma_{y^*})$, with
$\Sigma_{y^*}=(1-y^*) \gamma_1 + \gamma_2$.  We assume $\gamma_1$ is the identity matrix and $\gamma_2$ is the compound symmetric matrix, 

$\gamma_{2ij}=\begin{cases} 1, & \mbox{if } i\mbox{  $=j$} \\ \rho, & \mbox{if } i\mbox{ $= j+1  or  j-1$} \\
0, & \mbox{if } i \mbox { $ \neq j-1,j,j+1$} \end{cases}$, 

\noindent in which we call $\rho$ the `Effect Size'.
 Under the null, there is no change in the covariance structure,  i.e.  $\gamma_2$ is the identity matrix, as is $\Sigma_{y^*}$ for all $y^*$. As $\rho$ increases, the relationship between the covariance and $y^*$ becomes stronger.
Figure \ref{Fig:power_conty} shows that the power for the proposed statistics is near the intended
 $\alpha=0.05$ when $\rho=0$. Figure \ref{Fig:power_conty} also shows that the directional statistic $S$ is the most powerful approach overall. 

\subsubsection{Simulation Model 5.}
\begin{figure}[t]
	\begin{center}
		\includegraphics[width=8cm]{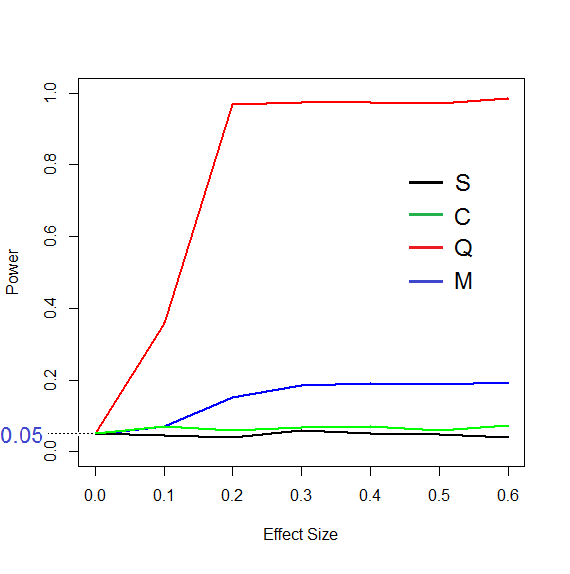}
		\caption{Power comparision among $S$, $Q$, $C$, $M$ for Simulation Model 5. The dashed line at $\alpha=0.05$ indicates that all the proposed methods control type I error well under the null ($\rho=0$). The effect size $\rho$ ranges from $0$ to $0.6$. $Q$ is the most powerful method among these statistics for this simulation model.}
		\label{Fig:power_conty2}
	\end{center}
\end{figure}

This simulation model is a bit more complex, following a similar approach used in \cite{cai2013two}.
The approach generates covariance matrices that are non-directional in relationship to $y$.
 and with no overall variation in magnitude, while respecting the need for positive definiteness. 
  To an initial $p\times p$ identity matrix $I$, $\Sigma^{*(1)}$ was formed by drawing the first $p/2 \times p/2$ off-diagonal elements from $U[-\rho,\rho]$, followed by $\Sigma^{*(2)}=\Sigma^{*(1)}+\Sigma^{{*(1)}T}$, and $\Sigma_1=\Sigma^{*(2)}+(\lambda_{min}(\Sigma^{*(2)})+0.05) I$. $\Sigma_2$ is formed by reversing the rows and columns of $\Sigma_1$, and finally $\Sigma_{y^*}=\Sigma_1 (1-y^*)+\Sigma_2 y^*$, where $y^*$ is the result of  linear rescaling of $y$ to the $[0,1]$ interval as in the previous subsection.
Here $\rho\in[0,1)$ serves as an effect size, and $\Sigma_1$ and $\Sigma_2$ differ in the groups of genes that show correlation structure, but otherwise are the same in the average magnitude of elements and show no directionality. Figure \ref{Fig:power_conty2} provides the power comparision among the four proposed methods. As expected, $S$ and $C$ have little or no power, while $M$ has extremely modest power.
The statistic $Q$ benefits from aggregation of covariance squared differences, and thus has much more power than the other methods.  All methods control type I error properly (dashed line at 0.05 in Figure \ref{Fig:power_conty2}).

\section{Analysis of two datasets}
We further illustrate the methods using two examples. The first is a two-sample analysis of a methylation dataset published by \cite{tsaprouni2014cigarette}. The data (GSE50660, Illumina Infinium HumanMethylation 450 BeadChip) consist of peripheral blood methylation signal from each of 464 individuals, with the binary experimental variable contrasting $179$ who never smoked ($y=0$) compared to $285$ former/current smokers ($y=1$).  The BeadStudio quantitative methylation signal was averaged for methylation sites within each of 22,003 genes using site-to-gene annotation, followed by annotation of genes to each of 4,512 Gene Ontology Biological Process terms, using Bioconductor v. 3.3. Covariates used were age, sex, and three significant surrogate variables determined using SVA (\cite{leek2012sva}).

The methylation signal for genes within each BP term were used as data matrices for each analysis, as a ``pathway" analysis for changes in covariance signal to be associated with smoking status. The data matrices were first residualized using the covariates and the effect of each row $x_{i.}$ on $y$, to identify BP terms that were specifically associated with changes in covariance rather than changes in means.  Analysis using $S$, $C$, and $M$ used the analytic approximations described here, and analysis for $Q$ was performed with 10,000 permutations for each BP term.  

Among the Biological Process terms for $Q$, GO:0061323 (15 genes, cell proliferation involved in heart morphogenesis) was significant at the 0.05 level ($p=1.0X10^{-5}$, FDR $q=0.045$). For this pathway, $C$ had $p=0.00015$, with the current/former smokers showing an overall higher magnitude of covariance than the never smokers. In addition, several pathways related to heart development were marginally significant with tied FDR $q=0.072$ (e.g., GO:0060914, 19 genes, heart).  These findings accord with findings by \cite{chowdhury2011maternal} that differentially methylated sites in a case-control study of maternal non-syndromic congenital heart defect-affected pregancies were enriched for genes involved in fetal development. Other statistics did not reach statistical significant after multiple testing. However, it is notable that the top pathway for $S$ was GO:0035094 (36 genes, response to nicotine, uncorrected $p=0.0004$).

%

To illustrate the utility of covariance testing in association with a continuous phenotype, we re-analyzed the data of \cite{van2002gene}, in which gene expression in breast tumors of 295 patients younger than 55 was examined for association with disease free survival.   We used martingale residuals, adjusted for age and sex and 50 surrogate variables, as a quantitative phenotype $y$.  Pathway analyses proceeded similarly as with the previous example, with Gene Ontology BP and KEGG pathways examined. $P$-values for $Q$ were determined by 100,000 permutations for high accuracy, and the remaining statistics used the asymptotic approximations.   
Of the four statistics proposed, three achieved false discovery $q<0.15$ for the most-significant gene set using Benjamini-Hochberg adjustment.
These were $Q$ (GO:0043254 regulation of protein complex assembly, 193 genes, $q=0.0066$), $C$ (GO:0022408 negative regulation of cell-cell adhesion, 75 genes, $q$=0.124), and $M$ (GO:1900221 regulation of beta-amyloid clearance, 5 genes, $q$=0.055). 

\begin{figure}[t]
	\begin{center}
		\includegraphics[width=8cm]{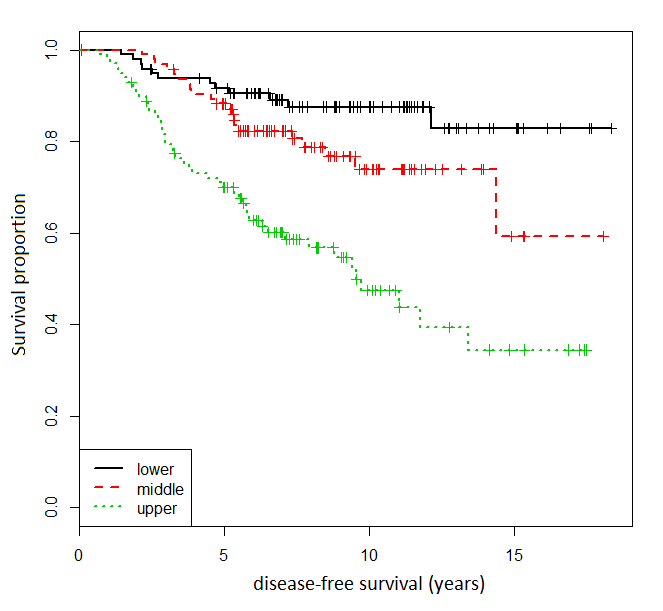}
		\caption{Kaplan-Meier curves for disease-free survival, breast cancer data of Van de Vijver et al. (2002) ($p=8\times 10^{-15}$). The curves correspond to tertiles of the risk scores for statistic $C$, for pathway GO:0022408 ``negative regulation of cell-cell adhesion."}
		\label{Fig:cellcell}
	\end{center}
\end{figure}

To further examine the finding for statistic $C$, we performed proportional hazards regression using the risk score vector $\{b_k\}$ as a predictor for disease-free survival while including the covariates described. The resulting Wald statistic for the risk scores was 
$p=8.0 \times 10^{-15}$. 
To visualize this striking result, we divided the risk score vector for $C$ into tertiles, and  
Figure \ref{Fig:cellcell} shows the corresponding Kaplan-Meier curves for disease-free survival. The result shows that high ``connectivity" of the genes in the pathway are associated with reduced disease-free survival, consistent with observations that loss of cellular adhesion promotes metastasis \citep{martin2013cancer}.

\section{Discussion}
We have proposed four covariance test statistics, in a straightforward trend-testing framework that applies to general $y$. 
The approach is not limited by $p$, $n$, or whether $y$ is discrete or continuous. The availability of a testing for continuous $y$ is a distinct advantage over previous methods, making covariance testing a simple approach that can be applied in a huge variety of settings.  We propose that the approach can be part of a standard testing toolkit, and used to evaluate, e.g. pathway associations in high-throughput data, or the statistical significance of network discoveries.

The software is available upon request from the author.

\section{Acknowledgments}
This work was supported by R21HG007840 and EPA STAR RD83574701.

\section{Appendix}
\subsection{Theorem 1}

\textbf{Theorem 1}. 
{\it Define $V_{k,n}=W_{k,n}Y_k$, $k=1,...,n$, and $S_n=\sum_k V_{k,n}$.
The Lindeberg condition for $\{U_{k,n}\}$ states that for any
$\varepsilon>0$, 
\[
\frac{1}{\sum_k{\rm var}(U_{k,n})}\sum_k E\bigl((U_{k,n}-E(U_{k,n}))^2 I(|U_{k,n}-E(U_{k,n})|>\varepsilon \sqrt{\sum_k{\rm var}(U_{k,n})}~\bigr)\longrightarrow 0
\]
as $n\rightarrow\infty$.
We assume $E(Y^4)<\infty$ and $E(W_{1,n}^4)<\infty$ for all $n$, and the Lindeberg condition holds for $\{W_{k,n}^2\}$ and $\{V_{k,n}\}$.

Define
\[
R=\frac{S_n/n-\bar{W}_n\bar{Y}}{(n-1)s_W s_Y}
\]
where $s_W$ and $s_Y$ are the sample standard deviations of $W_{.,n}$ and $Y$,
and $Z=\sqrt{n}R$.
Then, under $H_0$ that $X$ and $Y$ are independent, for any $t$, 
$P(Z\le t) \rightarrow \Phi(t)$ as $n\rightarrow \infty$.}

\noindent{\it Proof:}
Note that the theorem is stated somewhat more generally than required, allowing that the columns of $X$ be non-identically distributed. 
Without loss of generality we can assume $E(V_{k,n}=0$), and by the triangular array version of
the Lindeberg Central Limit Theorem (\cite{ferguson1996course}, pg. 27)  
$T_n=S_n/\sqrt{{\rm var}(S_n)}\underset{D}{\rightarrow} N(0,1)$. Asymptotic normality
for $U_n=\sum_k W_{k,n}^2$ follows similarly.
The central limit theorem for the correlation coefficient $R$ then follows from multivariate normality of $T_n$ and $U_n$, and other terms as described in \cite{ferguson1996course}, pg. 27, including rescaling $R$ by $\sqrt{n}$ to achieve limiting variance of 1.

\subsection{An example of establishing conditions for Theorem 1 for $S$}
Suppose each of the columns of $X$ is $p$-variate multivariate normal $MVN(0,\Sigma)$, where both $p$ and $\Sigma$ may depend on $n$. Define $A_{k,n}=\sum_i X_{ik}$, $\xi^2_n={\rm var}(A_{k,n})=E(A_{k,n}^2)$, and 
$W_{k,n}^2=A_{k,n}^4=Z_{k,n}^4\xi_n^4$, where $Z_{k,n}\sim N(0,1)$ (and we drop the $n$ subscript from $Z$).  As $k$ is arbitrary, Theorem 1 requires establishing Lindeberg conditions for $W_{1,n}^2$, for which the stronger Lyapunov condition is sufficient, and we compute
\[
L_n=\frac{1}{(n {\rm var}(W_{1,n}))^{1+\delta/2}}nE\bigl(|W_{1,n}^2-E(W_{1,n}^2)|^{2+\delta}\bigr)
=\frac{1}{n^{\delta/2} \xi_n^{8+4\delta}{\rm var}(Z^4)} E\bigl((\xi_n^4 |Z_1^4-E(Z_{1,n}^4)|)^{2+\delta}\bigr)
\]
for $\delta>0$, and the terms involving $\xi_n$ cancel while other terms do not depend on $n$.  It is thus clear that $L_n\rightarrow 0$ as $n\rightarrow \infty$, satisfying the condition. We apply the same approach to $\sum_k W_{k,n}Y_k$, similarly yielding cancellation of $\xi_n$ and $L_n\rightarrow 0$.

\subsection{Theorem 2}

\noindent \textbf{Theorem 2}.  Suppose that the correlation condition (C1), tail condition (C2) and the moment condition (C3) shown below hold. Then under $H_0$, for any $t$, 
\[
P(M-4\log(p)+\log(\log(p))\le t)\longrightarrow \exp\bigl(-\frac{1}{\sqrt{8\pi}}\exp(-t/2)\bigr),
\] 
as $n\longrightarrow \infty$.

\noindent{\it Proof:}

\noindent The proof follows from the results of Cai et al. (2013), who proved the result for comparison of sample covariances in two samples, and the result for our statistic $M$ nearly follows as a special case.
Thus we briefly point to the steps in their proof, and for clarity we restate the parameters and statistics using subscript notation and parameterization analogous to Cai et al. (2013), where $X_{ki}$ is the data value for feature $i$ and sample $k$, with mean $\mu_i$, realized value $x_{ki}$ and sample mean $\bar{x}_i$ across the $n$ samples. For a randomly drawn sample, the subscript $k$ may be dropped.
The experimental variable $Y$ has mean $\mu_Y$, realized value $y_k$ and sample mean $\bar{y}$.
First, we have $M_{ij}=(n-1) r^2_{ij}$, where $r$ is the Pearson correlation between $z_{ij.}$ and $y$.
Although the definition of our $M_{ij}$ is proportional to a squared correlation between vectors $z_{.ij}=(x_{.i}-\bar{x}_i)(x_{.j}-\bar{x}_j)$ and $y$, we note that, under the null hypothesis,
\[
{\rm cov}((X_i-\mu_i)(X_j-\mu_j),(Y-\mu_Y))={\rm cov}((X_i-\mu_i),(X_j-\mu_j)(Y-\mu_Y))=0,
\]
and both covariances include the same term $E((X_i-\mu_i)(X_j-\mu_j)(Y-\mu_Y))$.
Accordingly, it will be useful below to reflect the different grouping of terms as follows. Let
$W_{ki}=(X_{ki}-\mu_i)(Y_k-\mu_Y)$.
We define
the pair $\{U_{i[j]}, V_{j[i]}\}$, where $U_{i[j]}=(X_i-\mu_i), V_{j[i]}=W_j$ if $i\le j$,
and  $U_{i[j]}=W_i, V_{j[i]}=(X_j-\mu_j)$ if $i>j$.

The results of Cai et al. (2013) will be applied as tests of covariance between $U_{i[j]}$ and $V_{j[i]}$, compared to a null covariance of zero. Our main result is essentially a special case of Theorem 1 of Cai et al. (2013), as it may be viewed as a ``one sample" version of their two-sample problem, with some  modifications to account for our use of a different variance estimate.

We define $\sigma_{ij}=E\bigl(U_{i[j]} V_{j[i]}), \theta_{ij}={\rm var}\bigl(U_{i[j]} V_{j[i]}\bigr)$,  and under the null hypothesis $\sigma_{ij}=0$ for all $i\ne j$. We define
$\rho_{ij}=corr(X_i, X_j)$.
Our estimators are
$\hat\sigma_{ij}=\frac{1}{n} \sum_k \bigl( (x_{ki}-\bar{x}_i)(x_{kj}-\bar{x}_j)(y_k-\bar{y})\bigr)
=\frac{1}{n} \sum_k z_{kij}(y_k-\bar{y})\bigr)=\frac{1}{n} \sum_k u_{ki[j]}v_{kj[i]}$
and $\hat{\theta}_{ij}=\frac{n}{n-1}s_z^2 s_y^2$, where $s_z^2$ and $s_y^2$ are the sample variances of $z_{.ij}$ and $y$. Then for feature pair $\{i,j\}$, we define $ M_{ij}=\hat\sigma_{ij}^2/(\hat\theta_{ij}/n)=(n-1)r_{ij}^2$, where $r_{ij}$ is the sample Pearson correlation between $z_{.ij}$ and $y$, and our final statistic is $M=\max_{i,j} (M_{ij})$.

\bigskip
\noindent In Cai et al. (2013), several conditions (denoted C1-C3) were assumed to hold, and our analogues are required. 

\medskip
{\noindent \bf Condition (C1)} (Cai et al., 2013). For $r\in (0,1)$, define ${\boldsymbol\Lambda}(r)=\{1\le i\le p:|\rho_{ij}|>r$ for some $j\ne i$.
Suppose there exists a subset $\Upsilon\subset \{1,2,..,p\}$ with cardinality $o(p)$ and a constant $\alpha_0>0$ such that for all $\gamma>0,$ $\underset{1\le\j\le p,\j \notin \Upsilon}{\max}
s_j(\alpha_0)=o(p^\gamma)$ and there exists $r>1$ and a sequence $\Lambda_{p,r}$ such that the cardinality of 
${\boldsymbol\Lambda}(r) \le \Lambda_{p,r}=o(p)$.

\medskip
{\noindent \bf Condition (C2).} (Sub-Gaussian tail and polynomial tail condition, analogue of C2 and C2$^*$ in Cai et al. (2013)).  Suppose that $\log(p)=o(n^{1/5})$. There exist constants $\eta>0$ and $K>0$ such that
$ E(e^{\eta(X_i-\mu_i)^2/{{\rm var}(X_i)}})\le K, E(e^{\eta(W_i)^2/{\rm var}(W_i)})\le K$
for all $i$.  Alternatively, we assume that for some $\gamma_0, c_1 >0$, $p\le c_1 n^{\gamma_0}$ and $\epsilon>0$,
$ E(|(X_i-\mu_i|/{\rm var}(X_i)^{1/2})^{4\gamma_0+4+\epsilon} \le K$ and 
$ E(|(W_i-E(W_i))|/{\rm var}(W_i)^{1/2})^{4\gamma_0+4+\epsilon} \le K$
for all $i$.
Also, in either case we assume $\min_{1\le i\le j\le p} \theta_{ij}/({\rm var} (U_{i[j]}){\rm var}(V_{j[i]}))\ge \tau$
for some $\tau>0$. 

\medskip
{\noindent \bf Condition (C3).} (Analogue of C3 in Cai et al. (2013)).  
For any collection $i, j, k, l\in \{1,2,...,p\}$ we assume without loss of generality that
$i\le j$, $k\le l$, and we suppose there exists $\kappa \ge \frac{1}{3}$ such that  
\[
E\bigl(U_{i[j]}V_{j[i]}U_{k[l]}V_{l[k]} \bigr)=
\kappa \bigl(\sigma_{ij}\sigma_{kl}+E(U_{i[j]}V_{j[i]}) E(V_{j[i]}V_{l[k]}) +E(U_{i[j]}V_{l[k]}) E(V_{j[i]}U_{k[l]})\bigr).\]

We use Lemmas 1 and 2 from Cai et al. (2013), and the following lemma.

\medskip
\noindent {\it Analogue of Lemma 3 of Cai et al. (2013)}.
Under the conditions of C2, there exists some constant $C>0$ such that
\[
P\bigl(\max_{i,j} |\hat\theta_{ij}-\theta_{ij}|/({\rm var}(U_{i[j]}){\rm var}(V_{j[i]}))\bigr)\ge C\frac{ \varepsilon_n}{\log(p)}=O(p^{-1}+n^{-\epsilon/8}),
\]
where $\varepsilon_n=\max((\log(p)^{1/6}/n^{1/2},\log(p)^{-1})\rightarrow 0$ as $n, p\rightarrow\infty$.

\medskip
We have $\hat\theta_{ij}=s_{Z_{ij}}^2 s_Y^2.$  We first note that Lemma 3 from Cai et al. applies directly to $s_{Z_{ij}}^2$ as an estimator of ${\rm var}(Z_{ij})$.  Furthermore, $s_Y^2/\sigma_Y^2=O_P(n^{-1})$, and so 
\[
P\bigl(\max_{i,j} |s_{Z_{ij}^2}s_Y^2-\sigma_{Z_{ij}}^2\sigma_Y^2|)\ge C\frac{ \varepsilon_n}{\log(p)}=O(p^{-1}+n^{-\epsilon/8}),
\]
using variance scaling wlog as in the Supplemental Appendix of Cai et al. (2013).

\bigskip
With the above assumptions in hand, the proof follows from the proof of Theorem 1 in Cai et al. (2013), where in each instance out $\hat\sigma_{ij}$ is substituted for (say) $\hat\sigma_{ij1}$
from Cai et al., and zero substituted for $\hat\sigma_{ij2}$, and 
a single denominator $\hat\theta_{ij}/n$ in place of the Cai denominator. 
Condition C1 prevents excessive correlation of features, while C2 ensures that large deviations in the data do not prevent limiting convergence of extreme $\hat\sigma_{ij}$.
\clearpage
\bibliographystyle{biorefs}
\bibliography{yihui2}

\begin{thebibliography}{99}

\bibitem[Anderson(1962)Anderson]{anderson1962introduction}
\textsc{Anderson, Theodore~Wilbur}. (1962).
\newblock An introduction to multivariate statistical analysis.
\newblock {\em Technical Report}, Wiley New York.

\bibitem[Cai \emph{and others}(2013)Cai, Liu and Xia]{cai2013two}
\textsc{Cai, Tony, Liu, Weidong and Xia, Yin}. (2013).
\newblock Two-sample covariance matrix testing and support recovery in
  high-dimensional and sparse settings.
\newblock {\em Journal of the American Statistical
  Association\/}~\textbf{108}(501), 265--277.

\bibitem[Chowdhury \emph{and others}(2011)Chowdhury, Erickson, MacLeod, Cleves,
  Hu, Karim and Hobbs]{chowdhury2011maternal}
\textsc{Chowdhury, Shimul, Erickson, Stephen~W, MacLeod, Stewart~L, Cleves,
  Mario~A, Hu, Ping, Karim, Mohammad~A and Hobbs, Charlotte~A}. (2011).
\newblock Maternal genome-wide dna methylation patterns and congenital heart
  defects.
\newblock {\em PloS one\/}~\textbf{6}(1), e16506.

\bibitem[Ferguson(1996)Ferguson]{ferguson1996course}
\textsc{Ferguson, Thomas~Shelburne}. (1996).
\newblock {\em A course in large sample theory\/}, Volume~49. Chapman \& Hall
  London.

\bibitem[Golub and Van~Loan(2012)Golub and Van~Loan]{golub2012matrix}
\textsc{Golub, Gene~H and Van~Loan, Charles~F}. (2012).
\newblock {\em Matrix computations\/}, Volume~3. JHU Press.

\bibitem[Good(2002)Good]{good2002extensions}
\textsc{Good, Phillip~I}. (2002).
\newblock Extensions of the concept of exchangeability and their applications.
\newblock {\em Journal of Modern Applied Statistical Methods\/}~\textbf{1}(2),
  34.

\bibitem[Isogai(2016)Isogai]{isogai2016building}
\textsc{Isogai, Takashi}. (2016).
\newblock Building a dynamic correlation network for fat-tailed financial asset
  returns.
\newblock {\em Applied Network Science\/}~\textbf{1}(1), 7.

\bibitem[John(1971)John]{john1971some}
\textsc{John, S}. (1971).
\newblock Some optimal multivariate tests.
\newblock {\em Biometrika\/}~\textbf{58}(1), 123--127.

\bibitem[Leek \emph{and others}(2012)Leek, Johnson, Parker, Jaffe and
  Storey]{leek2012sva}
\textsc{Leek, Jeffrey~T, Johnson, W~Evan, Parker, Hilary~S, Jaffe, Andrew~E and
  Storey, John~D}. (2012).
\newblock The sva package for removing batch effects and other unwanted
  variation in high-throughput experiments.
\newblock {\em Bioinformatics\/}~\textbf{28}(6), 882--883.

\bibitem[Li \emph{and others}(2012)Li, Chen  et~al.]{li2012two}
\textsc{Li, Jun, Chen, Song~Xi  \emph{and others}}. (2012).
\newblock Two sample tests for high-dimensional covariance matrices.
\newblock {\em The Annals of Statistics\/}~\textbf{40}(2), 908--940.

\bibitem[Li(2002)Li]{li2002genome}
\textsc{Li, Ker-Chau}. (2002).
\newblock Genome-wide coexpression dynamics: theory and application.
\newblock {\em Proceedings of the National Academy of
  Sciences\/}~\textbf{99}(26), 16875--16880.

\bibitem[Martin \emph{and others}(2013)Martin, Ye, Sanders, Lane and
  Jiang]{martin2013cancer}
\textsc{Martin, Tracey~A, Ye, Lin, Sanders, Andrew~J, Lane, Jane and Jiang,
  Wen~G}. (2013).
\newblock Cancer invasion and metastasis: molecular and cellular perspective.

\bibitem[McKenzie \emph{and others}(2016)McKenzie, Katsyv, Song, Wang and
  Zhang]{mckenzie2016dgca}
\textsc{McKenzie, Andrew~T, Katsyv, Igor, Song, Won-Min, Wang, Minghui and
  Zhang, Bin}. (2016).
\newblock Dgca: A comprehensive r package for differential gene correlation
  analysis.
\newblock {\em BMC systems biology\/}~\textbf{10}(1), 106.

\bibitem[Modena \emph{and others}(2016)Modena, Kurian, Gaber, Waalen, Su,
  Gelbart, Mondala, Head, Papp, Heilman  et~al.]{modena2016gene}
\textsc{Modena, Brian~D, Kurian, Sunil~M, Gaber, Lillian~W, Waalen, Jill, Su,
  Andrew~I, Gelbart, Terri, Mondala, Tony~S, Head, Steven~R, Papp, Suzanne,
  Heilman, Raymond  \emph{and others}}. (2016).
\newblock Gene expression in biopsies of acute rejection and interstitial
  fibrosis/tubular atrophy reveals highly shared mechanisms that correlate with
  worse long-term outcomes.
\newblock {\em American Journal of Transplantation\/}~\textbf{16}(7),
  1982--1998.

\bibitem[Morgun \emph{and others}(2006)Morgun, Shulzhenko, Perez-Diez, Diniz,
  Sanson, Almeida, Matzinger and Gerbase-DeLima]{morgun2006molecular}
\textsc{Morgun, Andrey, Shulzhenko, Natalia, Perez-Diez, Ainhoa, Diniz,
  Rosiane~VZ, Sanson, Gerdine~F, Almeida, Dirceu~R, Matzinger, Polly and
  Gerbase-DeLima, Maria}. (2006).
\newblock Molecular profiling improves diagnoses of rejection and infection in
  transplanted organs.
\newblock {\em Circulation Research\/}~\textbf{98}(12), e74--e83.

\bibitem[Pesarin and Salmaso(2010)Pesarin and Salmaso]{pesarin2010permutation}
\textsc{Pesarin, Fortunato and Salmaso, Luigi}. (2010).
\newblock {\em Permutation tests for complex data: theory, applications and
  software\/}. John Wiley \& Sons.

\bibitem[Sandholm \emph{and others}(2012)Sandholm, Salem, McKnight, Brennan,
  Forsblom, Isakova, McKay, Williams, Sadlier, M{\"a}kinen
  et~al.]{sandholm2012new}
\textsc{Sandholm, Niina, Salem, Rany~M, McKnight, Amy~Jayne, Brennan, Eoin~P,
  Forsblom, Carol, Isakova, Tamara, McKay, Gareth~J, Williams, Winfred~W,
  Sadlier, Denise~M, M{\"a}kinen, Ville-Petteri  \emph{and others}}. (2012).
\newblock New susceptibility loci associated with kidney disease in type 1
  diabetes.
\newblock {\em PLoS genetics\/}~\textbf{8}(9), e1002921.

\bibitem[Tsaprouni \emph{and others}(2014)Tsaprouni, Yang, Bell, Dick, Kanoni,
  Nisbet, Vi{\~n}uela, Grundberg, Nelson, Meduri
  et~al.]{tsaprouni2014cigarette}
\textsc{Tsaprouni, Loukia~G, Yang, Tsun-Po, Bell, Jordana, Dick, Katherine~J,
  Kanoni, Stavroula, Nisbet, James, Vi{\~n}uela, Ana, Grundberg, Elin, Nelson,
  Christopher~P, Meduri, Eshwar  \emph{and others}}. (2014).
\newblock Cigarette smoking reduces dna methylation levels at multiple genomic
  loci but the effect is partially reversible upon cessation.
\newblock {\em Epigenetics\/}~\textbf{9}(10), 1382--1396.

\bibitem[Van De~Vijver \emph{and others}(2002)Van De~Vijver, He, Van't~Veer,
  Dai, Hart, Voskuil, Schreiber, Peterse, Roberts, Marton  et~al.]{van2002gene}
\textsc{Van De~Vijver, Marc~J, He, Yudong~D, Van't~Veer, Laura~J, Dai, Hongyue,
  Hart, Augustinus~AM, Voskuil, Dorien~W, Schreiber, George~J, Peterse,
  Johannes~L, Roberts, Chris, Marton, Matthew~J  \emph{and others}}. (2002).
\newblock A gene-expression signature as a predictor of survival in breast
  cancer.
\newblock {\em New England Journal of Medicine\/}~\textbf{347}(25), 1999--2009.

\bibitem[Yuan \emph{and others}(2016)Yuan, Deng, Tang and Li]{yuan2016network}
\textsc{Yuan, Huili, Deng, Minghua, Tang, Nelson~LS and Li, Zhenye}. (2016).
\newblock A network based covariance test for detecting multivariate eqtl in
  saccharomyces cerevisiae.
\newblock In:  {\em BMC systems biology\/}, Volume~10. BioMed Central. p.~S8.

\bibitem[Zeng \emph{and others}(2014)Zeng, Miyazawa, Kloepfer and
  Harris]{zeng2014deletion}
\textsc{Zeng, Fenghua, Miyazawa, Tomoki, Kloepfer, Lance~A and Harris,
  Raymond~C}. (2014).
\newblock Deletion of erbb4 accelerates polycystic kidney disease progression
  in cpk mice.
\newblock {\em Kidney international\/}~\textbf{86}(3), 538.

\bibitem[Zhou \emph{and others}(2013)Zhou, Mayhew, Sun, Xu, Zou and
  Wright]{zhou2013space}
\textsc{Zhou, Yi-Hui, Mayhew, Gregory, Sun, Zhibin, Xu, Xiaolin, Zou, Fei and
  Wright, Fred~A}. (2013).
\newblock Space--time clustering and the permutation moments of quadratic
  forms.
\newblock {\em Stat\/}~\textbf{2}(1), 292--302.

\bibitem[Zhou and Wright(2015)Zhou and Wright]{zhou2015hypothesis}
\textsc{Zhou, Yi-Hui and Wright, Fred~A}. (2015).
\newblock Hypothesis testing at the extremes: fast and robust association for
  high-throughput data.
\newblock {\em Biostatistics\/}, kxv007.

\end{thebibliography}

\clearpage

\begin{table}[h!]
  \centering
  \caption{Type I error comparison, Sim. Model 1, $X_k \sim N(0,\Sigma_k)$, $\Sigma_1=\Sigma_2$}
  \label{tab:typeInormal}
  \begin{tabular}{c|c|cccccc}
\hline
\hline
    $n_1=n_2$ & Method &  p=32 & p=64 & p=128  & p=256  & p=512  & p=700\\
\hline
20&$S$ &0.041&0.041&0.047& 0.047& 0.041& 0.042\\
& $Q$ &0.055&0.058&0.046& 0.043& 0.043& 0.052\\
&$C$ &0.052&0.050&0.049& 0.051& 0.048& 0.050\\
& $Li-Chen$ &0.044&0.054&0.051& 0.048& 0.051& 0.038\\
& $Cai$ & 0.092&0.14&0.139&0.204&0.211&0.263\\
& $M$ & 0.053& 0.054 & 0.050 & 0.052 &0.051 &0.050\\
\hline
50& $S$ &0.047 &0.043& 0.050& 0.043& 0.047& 0.047\\
&$Q$ &0.052 &0.041& 0.045& 0.049& 0.042& 0.046\\
&$C$ &0.045 &0.050& 0.052& 0.055& 0.046& 0.046\\
&$Li-Chen$ &0.052 &0.060& 0.033& 0.043& 0.054& 0.049\\
&$Cai$&0.042&0.068&	0.058&0.065&0.055&0.059\\
& $M$ & 0.059 & 0.054& 0.051 & 0.048 & 0.051&0.050\\
\hline
80 & $S$ & 0.047 &0.064& 0.049& 0.049& 0.047& 0.043\\
& $Q$ & 0.065& 0.051& 0.040& 0.046& 0.044 &0.048\\
& $C$ & 0.051 &0.049& 0.051 &0.052& 0.047 &0.047\\
& $Li-Chen$ & 0.054 &0.060& 0.047& 0.048& 0.052& 0.053\\
& $Cai$ & 0.052	&0.056&	0.043&	0.052&	0.058&	0.041\\
& $M$ & 0.046 & 0.046 & 0.051 &0.047 & 0.050&0.049\\
\hline
100 & $S$ & 0.051& 0.051& 0.050& 0.048& 0.049& 0.050\\
& $Q$ & 0.039& 0.051& 0.050& 0.040& 0.060& 0.053\\
& $C$ & 0.050& 0.050&0.047& 0.047& 0.043& 0.053\\
& $Li-Chen$ &0.056& 0.049& 0.052& 0.046& 0.049& 0.048\\
& $Cai$ &0.050&0.052&0.043&0.039&0.036&0.047 \\
& $M$ & 0.047 & 0.054 & 0.048 & 0.050&0.048& 0.046\\
\hline
\hline 
  \end{tabular}
\end{table}

\begin{table}[h!]
  \centering
  \caption{Type I error comparison, Sim. Model 2, $\Sigma_1=\Sigma_2$, elements following centered gamma}
  \label{tab:typeIgamma}
  \begin{tabular}{c|c|cccccc}
\hline
\hline
    $n_1=n_2$ & Method &  p=32 & p=64 & p=128  & p=256  & p=512  & p=700\\
\hline
20 & $S$ & 0.042& 0.041& 0.042& 0.042& 0.046& 0.044\\
& $Q$ &0.057 &0.046& 0.066& 0.050& 0.047& 0.042\\
& $C$ & 0.040 &0.041& 0.042& 0.045& 0.046& 0.046 \\
& $Li-Chen$ & 0.158 &0.112&0.083&0.071&0.053&0.063\\
& $Cai$& 0.048&0.048	&0.058	&0.055	&0.083	&0.085\\
& $M$ &0.039&0.039&0.050&0.037&0.042&0.046\\
\hline
50& $S$&0.040 &0.043& 0.046 &0.050& 0.049& 0.049\\
&$Q$&0.055& 0.049& 0.055& 0.043& 0.051 &0.042\\
&$C$&0.050 &0.043 &0.046 &0.049& 0.040 &0.049\\
&$Li-Chen$&0.048	&0.048	&0.058&	0.055	&0.083	&0.085\\
& $Cai$ &0.016	&0.013&	0.010&	0.007	&0.003&	0.004\\
& $M$ &0.051 & 0.05& 0.055 & 0.042&0.050 &0.048\\
\hline
80& $S$ &0.048 &0.044 &0.040 &0.045 &0.044 &0.051\\
& $Q$ &0.056 &0.048 &0.043 &0.042 &0.038 &0.057\\
& $C$ & 0.040 &0.046 &0.044 &0.043 &0.050 &0.045\\
& $Li-Chen$ & 0.165&0.141&0.090&0.059&0.051&0.056\\
&$Cai$ & 0.019&0.010&	0.005&	0.006&	0.005&	0.002\\
& $M$ & 0.059 & 0.039 & 0.045& 0.045& 0.048&0.051 \\
\hline
100& $S$ & 0.044 &0.048 &0.048 &0.051 &0.053& 0.050\\
& $Q$ & 0.045 &0.042& 0.049& 0.049& 0.051& 0.039\\
& $C$ & 0.049 &0.051& 0.045 &0.043 &0.052& 0.053\\
& $Li-Chen$ & 0.176&0.133&0.088&0.069	&0.050&	0.046\\
& $Cai$ &0.013&0.009&	0.007	&0.003	&0.003&	0.003\\
& $M$ & 0.059 & 0.041& 0.058 & 0.052&0.053&0.050\\
\hline
\hline 
  \end{tabular}
\end{table}

\begin{table}[h!]
  \centering
  \caption{Power comparison, Sim. Model 3, $X_k \sim N(0,\Sigma_k)$, $\Sigma_1\ne\Sigma_2$}
  \label{tab:powernormal}
  \begin{tabular}{c|c|cccccc}
\hline
\hline
    $n_1=n_2$ & Method &  p=32 & p=64 & p=128  & p=256  & p=512  & p=700\\
\hline
20 & $S$ & 0.173 & 0.180& 0.187 &0.184& 0.179& 0.177\\
&$Q$ &0.211 &0.231 &0.235& 0.234 &0.221 &0.213\\
&$C$ &0.632 &0.855 &0.977& 1.000&   1.000    &1.000\\
&$Li-Chen$ &0.273	&0.273	&0.252	&0.285	&0.269	&0.272\\
& $Cai$ & 0.138	&0.140	&0.164	&0.204	&0.233	&0.282\\
& $M$ & 0.129 & 0.072 & 0.050& 0.061 & 0.083 & 0.054\\
\hline
50 & $S$ &  0.450 &0.462& 0.480& 0.468& 0.481& 0.479\\
& $Q$ & 0.705 &0.751& 0.803 &0.809& 0.772& 0.789\\
& $C$ & 0.989 &1.000&1.000 &1.000& 1.000& 1.000\\
& $Li-Chen$ & 0.752	&0.800&	0.824&0.861&	0.839&	0.857\\
& $Cai$ & 0.234&0.163&0.146&0.136&0.104&0.084\\
& $M$ & 0.270 & 0.133& 0.092 & 0.122 & 0.034& 0.051\\
\hline 
80 & $ S$ & 0.678 &0.685& 0.696& 0.708& 0.703& 0.702\\
&$Q$&0.955 &0.972 &0.991 &0.995 &0.992 &0.992\\
&$C$&1.000 &1.000 &1.000 &1.000   & 1.000    &1.000\\
& $Li-Chen$ &0.941&0.980&0.992	&0.994	&0.996	&0.998\\
& $Cai$ &0.496	&0.420&	0.377	&0.316&	0.246	&0.189\\
& $M$ & 0.574 & 0.394 & 0.333& 0.242& 0.253 & 0.201\\
\hline
100& $S$& 0.775 &0.790 &0.797&0.798 &0.780 &0.809\\
& $Q$ & 0.991 &0.997 &0.999& 1.000& 1.000& 1.000\\
& $C$ &1.000 &1.000 &1.000 &1.000 &   1.000   & 1.000\\
& $Li-Chen$ &0.997	&1.000&	0.999&	1.000&	1.000&	1.000\\
& $Cai$ &0.700	&0.652	&0.557&	0.508&	0.423	&0.406\\
& $M$ & 0.700 & 0.649 & 0.601 & 0.487 &0.375 & 0.374\\
\hline
\hline
  \end{tabular}
\end{table}

\end{document}